\documentclass{PoS}

\usepackage{bm}
\newcommand{\be}{\begin{equation}}
\newcommand{\ee}{\end{equation}}
\newcommand{\bea}{\begin{eqnarray}}
\newcommand{\eea}{\end{eqnarray}}

\newcommand{\pup}{p^\uparrow}

\newcommand{\qup}{q^\uparrow}

\def\lsim{\mathrel{\rlap{\lower4pt\hbox{\hskip1pt$\sim$}}\raise1pt\hbox{$<$}}}
\def\gsim{\mathrel{\rlap{\lower4pt\hbox{\hskip1pt$\sim$}}\raise1pt\hbox{$>$}}}
\def\nostrocostruttino#1\over#2{\mathrel{\mathop{\kern 0pt \rlap
{\hbox{$#1$}}} \hbox{\kern-.135em $#2$}}}

\title{Quasi-real photon contribution to $A_N$ in $\ell p^\uparrow \to \pi\,$ X within a TMD approach}

\ShortTitle{Quasi-real photon contribution to $A_N$}

\author{\speaker{Umberto D'Alesio}
\\
        Dipartimento di Fisica, Universit\`a di Cagliari, Cittadella Universitaria, and \\
INFN, Sezione di Cagliari, I-09042 Monserrato (CA), Italy\\
        E-mail: \email{umberto.dalesio@ca.infn.it}}

\author{Carlo Flore\\
Dipartimento di Fisica, Universit\`a di Cagliari, Cittadella Universitaria, and \\
INFN, Sezione di Cagliari, I-09042 Monserrato (CA), Italy\\
        E-mail: \email{carlo.flore@ca.infn.it}}

\author{Francesco Murgia\\
INFN, Sezione di Cagliari, I-09042 Monserrato (CA), Italy\\
        E-mail: \email{francesco.murgia@ca.infn.it}}

\abstract{Within a TMD approach we discuss the impact of quasi-real (Weizs\"acker-Williams) photon contribution to the transverse single-spin asymmetry, $A_N$, for the inclusive process $\ell p^\uparrow \to \pi\, X$. This study extends a previous one where only the leading-order contribution was taken into account. Predictions are obtained adopting the Sivers and transversity distributions and the Collins fragmentation functions as extracted from fits to the azimuthal asymmetries measured in semi-inclusive deep inelastic scattering and $e^+e^-$ annihilation processes. The description of the available data is very good, showing a clear general improvement with respect to the previous analysis (already quite promising). This result represents a further step towards a unified TMD treatment of SSAs in $\ell\,p^\uparrow \to\ell'\, h\, X$ and $\ell \,p^\uparrow \to h\, X$ processes.
}

\FullConference{QCD Evolution 2016\\
		May 30-June 03, 2016\\
		National Institute for Subatomic Physics (Nikhef), Amsterdam}

\begin{document}

\section{Introduction}

The issue of the validity of factorization in terms of transverse momentum dependent distributions (TMDs) for inclusive processes where only one large energy scale is detected is challenging and still under debate. The description within a TMD approach (as well as in a twist-three formalism) of the large data sets for $A_N$ measured in inclusive pion production in $p^\uparrow p$ collisions is the most striking example (see for instance Ref.~\cite{D'Alesio:2007jt}).

In Refs.~\cite{Anselmino:2009pn,Anselmino:2014eza} this issue has been investigated in a somehow simpler inclusive process, still with a single large energy scale, but very close to the semi-inclusive deep inelastic scattering (SIDIS) process, for which TMD factorization has been proven~\cite{Collins:2011zzd,GarciaEchevarria:2011rb,Echevarria:2014rua}.
We refer to transverse single-spin asymmetries (SSAs) for the $\ell \, \pup \to h \, X$ process, with the detection, in the lepton-proton center of mass ({\it c.m.}) frame, of a single large-$P_T$ final particle, typically a pion. The same process was also considered in Refs.~\cite{Koike:2002gm,Gamberg:2014eia} in the framework of collinear factorization with twist-three correlation functions.

These SSAs were computed assuming the TMD factorization at leading order (LO) and using the relevant TMDs (Sivers and Collins functions) as extracted from SIDIS and $e^+ e^-$ data.
In particular, in Ref.~\cite{Anselmino:2014eza} the theoretical estimates were compared with a selection\footnote{Only data for inclusive events in the backward target hemisphere and tagged events were considered.} of the experimental results by
the HERMES Collaboration~\cite{Airapetian:2013bim}, showing a good agreement in sign and size.

Here we want to extend this LO study including the contribution from quasi-real (Weizs\"acker-Williams) photon exchange, relevant in the kinematical configuration dominated by small $Q^2$ (for a complete and comprehensive study see Ref.~\cite{DAlesio:2017nrd}). This will allow us to improve the description of the fully-inclusive data and consider, for the first time, the anti-tagged data set, dominated by events in which the final lepton has a very small scattering angle. Notice that this data category was not included in the previous analysis because a simple LO approach (namely via $q\ell\to q \ell$) is expected to be not adequate.

\section{Formalism}
In Refs.~\cite{Anselmino:2009pn,Anselmino:2014eza}, to which we refer the reader for all details, a study of SSAs in $\ell p^\uparrow \to \pi \,X$ processes within a TMD approach was presented, restricting to a leading-order approximation.
Here, still within the same approach, we want to consider the impact of quasi-real photon exchange. As pointed out in Ref.~\cite{Hinderer:2015hra} this contribution, at least in a collinear approach and for unpolarized cross sections, represents in most cases only a small part of the next-to-leading order (NLO) calculations. On the other hand in a TMD scheme, as for the twist-three approach, NLO corrections are not available for  such a process and it is then worth to see to what extent the quasi-real photon exchange could play a role in the computation of spin asymmetries. On top of that, by including transverse momentum effects the estimates of unpolarized cross sections are enhanced w.r.t.~those computed in a collinear framework.

We consider the transverse single-spin asymmetry, $A_N$, for the process $\pup \ell \to h \, X$ in the proton-lepton {\it c.m.}
frame (with the polarized proton moving along the positive $Z_{\rm cm}$ axis)
 \be
A_N = \frac{d\sigma^\uparrow(\bm{P}_T) - d\sigma^\downarrow(\bm{P}_T)}
           {d\sigma^\uparrow(\bm{P}_T) + d\sigma^\downarrow(\bm{P}_T)}
    = \frac{d\sigma^\uparrow(\bm{P}_T) - d\sigma^\uparrow(-\bm{P}_T)}
           {2 \, d\sigma^{\rm unp}(\bm{P}_T)} \,, \label{an}
\ee
where $\bm{P}_T$ is the transverse momentum of the final hadron.
Notice that for a generic transverse polarization, $\bm{S}_T$, along an azimuthal direction $\phi_S$ in the chosen reference frame, in which the $\uparrow$ direction is given by $\phi_S = \pi/2$, one has:
\be
A(\phi_S, S_T) = \bm{S}_T \cdot (\hat{\bm{p}} \times \hat{\bm{P}}_T) \, A_N = S_T \sin\phi_S \, A_N \>, \label{phis}
\ee
where $\bm{p}$ is the proton momentum.

Assuming the validity of the TMD factorization scheme for the process $\pup \,\ell \to h \, X$, in which $P_T=|\bm{P}_T|$ is the only large scale detected, as discussed in Refs.~\cite{Anselmino:2009pn,Anselmino:2014eza}, in a leading-order approximation the main contribution to $A_N$ comes from the Sivers~\cite{Sivers:1989cc,Sivers:1990fh} and Collins~\cite{Collins:1992kk} effects and one has~\cite{D'Alesio:2004up,Anselmino:2005sh,D'Alesio:2007jt}:
\be
A_N =
\frac
{{\displaystyle \sum_{q} \int \frac{dx \, dz}
{16\,\pi^2 x\,z^2 s}}\;
d^2 \bm{k}_{\perp} \, d^3 \bm{p}_{\perp}\,
\delta(\bm{p}_{\perp} \cdot \hat{\bm{p}}'_q) \, J(p_\perp)
\> \delta(\hat s + \hat t + \hat u)
\> [\Sigma(\uparrow) - \Sigma(\downarrow)]^{q \ell \to q \ell}}
{{\displaystyle \sum_{q} \int \frac{dx \, dz}
{16\,\pi^2 x\,z^2 s}}\;
d^2 \bm{k}_{\perp} \, d^3 \bm{p}_{\perp}\,
\delta(\bm{p}_{\perp} \cdot \hat{\bm{p}}'_q) \, J(p_\perp)
\> \delta(\hat s + \hat t + \hat u)
\> [\Sigma(\uparrow) + \Sigma(\downarrow)]^{q \ell \to q \ell}} \>,
\label{anh}
\ee
with (dropping negligible contributions from other TMDs)
\bea
\,[\Sigma(\uparrow) - \Sigma(\downarrow)]^{q
\ell \to q \ell} &=& \frac{1}{2} \, \Delta^N\! f_{q/\pup}
(x,k_{\perp}) \cos\phi \, \left[\,|{\hat M}_1^0|^2 + |{\hat
M}_2^0|^2 \right] \,
D_{h/q} (z, p_{\perp})  \nonumber \\
&+& h_{1q}(x,k_{\perp}) \, \hat M_1^0 \hat M_2^0 \, \Delta^N\!
D_{h/\qup} (z, p_{\perp}) \, \cos(\phi' + \phi_q^h) \label{ds1}
\eea
and
\be
\,[\Sigma(\uparrow) +
\Sigma(\downarrow)]^{q \ell \to q \ell} =
f_{q/p} (x,k_{\perp}) \,
\left[\,|{\hat M}_1^0|^2 + |{\hat M}_2^0|^2 \right] \,
D_{h/q} (z, p_{\perp}) \>. \label{ss1}
\ee

All functions and all kinematical and dynamical variables appearing in the above equations are exactly defined in Ref.~\cite{Anselmino:2009pn} and its Appendices and in Ref.~\cite{Anselmino:2005sh}.

In order to include also possible contributions from quasi-real photon exchange we rely on the well-known Weizs\"acker-Williams (WW) approximation. Without entering into many details we only recall that in such an approximation the incoming lepton is considered as a source of real photons (with their proper distribution), which then enter the hard scattering process. In other words, we adopt the following factorization formula for the WW contribution to the process $\ell p \to h\, X$:
\be
\sigma^{\rm WW}(\ell p \to h\, X) = \int d y f_{\gamma/\ell}(y)\, \sigma(\gamma p\to h\, X) \,,
\ee
where $f_{\gamma/\ell}(y)$ is the number density of photons inside the lepton, carrying a lepton-momentum fraction $y$ ($p_\gamma = y p_\ell$) and $\sigma(\gamma p\to h \,X)$ is the cross section for the process $\gamma p\to h \,X$ initiated by a real photon. In particular, following Ref.~\cite{Hinderer:2015hra},  we adopt the expression
\be
f_{\gamma/\ell}(y) = \frac{\alpha}{2\pi} \frac{1+(1-y)^2}{y}\Biggl[\ln \Biggl({\frac{\mu^2}{y^2 m^2_{\ell}}}\Biggr) - 1 \Biggr] + \mathcal{O}(\alpha^2)
  \,,
\label{ww}
\ee
where $\alpha$ is the electromagnetic coupling constant, $\mu$ the factorization scale (set in the following equal to $P_T$) and $m_\ell$ the lepton mass.
Once again for a generic polarized process adopting the helicity formalism one can calculate all possible contributions in terms of TMDs (see Ref.~\cite{DAlesio:2017nrd} for all details). Here we simply remark that even for the real-photon initiated process only the Sivers and the Collins effects could give potentially sizeable contributions to the numerator of $A_N$, while once again only the unpolarized TMDs play a role in the denominator. At variance with the leading-order analysis, where there is only one partonic channel, namely $q\ell\to q\ell$, here we have to consider the following channels: $q\gamma\to q g$ and $g \gamma\to q\bar q$. This means that when we refer to the Sivers effect also the contribution from the gluon Sivers function has to be taken into account.

We can then rewrite the asymmetry under consideration in the following way:
\be
\label{ANWW}
A_N = \frac{d\Delta\sigma^{\rm LO} + d\Delta\sigma^{\rm WW}}{2[ d\sigma^{\rm LO} + d\sigma^{\rm WW}]}\,,
\ee
where
$d\Delta\sigma^{\rm LO}$ and $[2\,d\sigma^{\rm LO}]$ are nothing else than the numerator and the denominator in Eq.~(\ref{anh}), while for the WW pieces one has to make the following replacement:
\bea
 [\Sigma(\uparrow) \pm \Sigma(\downarrow)]^{q \ell \to q \ell} & \to &
  [\Sigma(\uparrow) \pm \Sigma(\downarrow)]^{q \gamma\to q g} + [\Sigma(\uparrow) \pm \Sigma(\downarrow)]^{q\gamma \to g q} \nonumber\\
 &+&[\Sigma(\uparrow) \pm \Sigma(\downarrow)]^{g \gamma \to q \bar q} + [\Sigma(\uparrow) \pm \Sigma(\downarrow)]^{g \gamma \to \bar q q}\label{sww}
\eea
and perform a further convolution over $y$ via $\int dy/y$. More explicitly, we have
\bea
\,[\Sigma(\uparrow) - \Sigma(\downarrow)]^{q \gamma \to q g}
& = & f_{\gamma/\ell}(y)
\Big\{\frac{1}{2} \, \Delta^N\! f_{q/\pup}(x,k_{\perp}) \cos\phi \, \left[\,|{\hat M}_1^0|^2 + |{\hat M}_2^0|^2 \right]^{q\gamma} \,D_{h/q} (z, p_{\perp}) \nonumber \\
&+&  h_{1q}(x,k_{\perp}) \, [\hat M_1^0 \hat M_2^0]^{q\gamma} \, \Delta^N\!D_{h/\qup} (z, p_{\perp}) \, \cos(\phi' + \phi_q^h) \Big\}\label{sww1a}\\
\,[\Sigma(\uparrow) +\Sigma(\downarrow)]^{q \gamma \to q g} & = & f_{\gamma/\ell}(y)
f_{q/p} (x,k_{\perp}) \,\left[\,|{\hat M}_1^0|^2 + |{\hat M}_2^0|^2 \right]^{q\gamma} \,
D_{h/q} (z, p_{\perp}) \label{sww1b}\\
\,[\Sigma(\uparrow) - \Sigma(\downarrow)]^{q \gamma \to g q } & = & \frac{1}{2}\, f_{\gamma/\ell}(y)
 \, \Delta^N\! f_{q/\pup}(x,k_{\perp}) \cos\phi \, \left[\,|{\hat M}_1^0|^2 + |{\hat M}_3^0|^2 \right]^{q\gamma} \,D_{h/g} (z, p_{\perp}) \label{sww2a}\\
\,[\Sigma(\uparrow) +\Sigma(\downarrow)]^{q \gamma \to g q} & = & f_{\gamma/\ell}(y)
f_{q/p} (x,k_{\perp}) \,\left[\,|{\hat M}_1^0|^2 + |{\hat M}_3^0|^2 \right]^{q\gamma} \,
D_{h/g} (z, p_{\perp}) \label{sww2b}\\
\,[\Sigma(\uparrow) - \Sigma(\downarrow)]^{g \gamma \to q \bar q} & = & \frac{1}{2}\,f_{\gamma/\ell}(y)
\, \Delta^N\! f_{g/\pup}(x,k_{\perp}) \cos\phi \, \left[\,|{\hat M}_2^0|^2 + |{\hat M}_3^0|^2 \right]^{g\gamma} \,D_{h/q} (z, p_{\perp})\label{sww3a}\\
\,[\Sigma(\uparrow) +\Sigma(\downarrow)]^{g \gamma \to q\bar q} & = & f_{\gamma/\ell}(y)
f_{g/p} (x,k_{\perp}) \,\left[\,|{\hat M}_2^0|^2 + |{\hat M}_3^0|^2 \right]^{g\gamma} \,
D_{h/q} (z, p_{\perp}) \>, \label{sww3b}
\eea
where, besides a common overall factor equal to $16\pi^2\alpha\alpha_s$,
\be
\left[\,|{\hat M}_1^0|^2 + |{\hat M}_2^0|^2 \right]^{q\gamma} =  \frac{16}{3}  e_q^2 \frac{\hat s^2+\hat u^2}{-\hat s \,\hat u}\,,\;\;
\left[\hat M_1^0 \hat M_2^0\right]^{q\gamma} = \frac{16}{3} e_q^2\,,\;\;
\left[\,|{\hat M}_1^0|^2 + |{\hat M}_3^0|^2 \right]^{q\gamma}  =  \frac{16}{3}  e_q^2 \frac{\hat s^2+\hat t^2}{-\hat s\, \hat t}
\label{M02}
\ee
\be
\left[\,|{\hat M}_2^0|^2 + |{\hat M}_3^0|^2 \right]^{g\gamma} =  2\,e_q^2\,\frac{\hat u^2 +\hat t^2}{\hat u\,\hat t}\,.
\ee
Notice that the Mandelstam invariants for the photon-parton subprocess have to be defined using $p_\gamma = yp_\ell$ and that for the process $g\gamma\to \bar q q$ one has the same expressions as in Eqs.~(\ref{sww3a}) and (\ref{sww3b}) with $D_{h/q}$ replaced by $D_{h/\bar q}$.

Some comments are in order: the WW contributions are of order $\alpha\alpha_s$; the structures are very similar to those appearing in the LO case; the Collins effect enters only in the channel $q\gamma \to q g$; the gluon Sivers effect appears in the channels $g\gamma \to q\bar q$ and $g \gamma \to \bar q q$.

\section{Results}

In this Section we show a selection of our results both for the unpolarized cross sections and the SSAs, with special attention to HERMES kinematics. For a more comprehensive study, including also predictions for ongoing and future experiments we refer the reader to Ref.~\cite{DAlesio:2017nrd}.

In our computation of the SSAs, based on the TMD factorization, we consider two different sets of Sivers and Collins functions (the latter coupled to the transversity distribution), as previously obtained in a series of papers from fits of SIDIS and $e^+e^-$
data~\cite{Anselmino:2005ea,Anselmino:2007fs,Anselmino:2008sga,Anselmino:2008jk}.

These sets, besides some different initial assumptions, differ in the choice of the collinear fragmentation functions (FFs). More precisely, for the fits~\cite{Anselmino:2005ea} and~\cite{Anselmino:2007fs} (SIDIS~1 set) we adopt the Kretzer set for the collinear FFs~\cite{Kretzer:2000yf}. For the fits~\cite{Anselmino:2008sga} and ~\cite{Anselmino:2008jk} (SIDIS~2 set) we adopt another set of FFs, namely the one by de Florian, Sassot and Stratmann (DSS)~\cite{deFlorian:2007aj}. The SIDIS~1 and SIDIS~2 sets are well representative of the extractions and their uncertainties. Concerning the gluon Sivers function we adopt the recent extraction of Ref.~\cite{D'Alesio:2015uta} (notice that we have a corresponding gluon Sivers function associated to each SIDIS set).

Let us start with the unpolarized cross sections for HERMES set-up, where the incoming lepton moves along the $+Z_{\rm cm}$ axis in the lepton-proton center of mass frame. This means that, defining as usual $x_F=2P_L/\sqrt s$ (with $P_L$ the longitudinal momentum of the final hadron), positive(negative) values of $x_F$ correspond to the backward(forward) proton hemisphere.

\begin{figure}[h!]
 \centering
 \includegraphics[scale=.65]{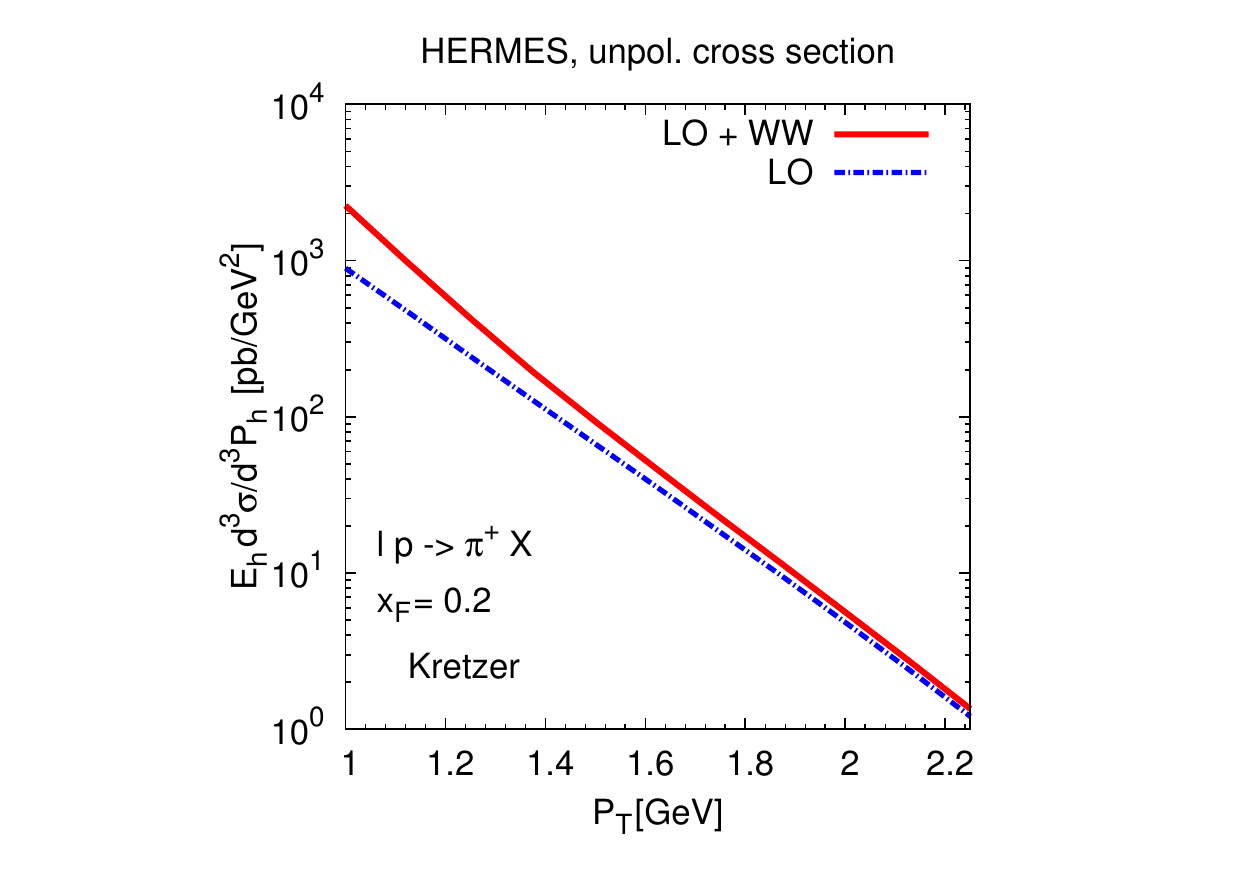}\hspace*{-2cm}
 \includegraphics[scale=.65]{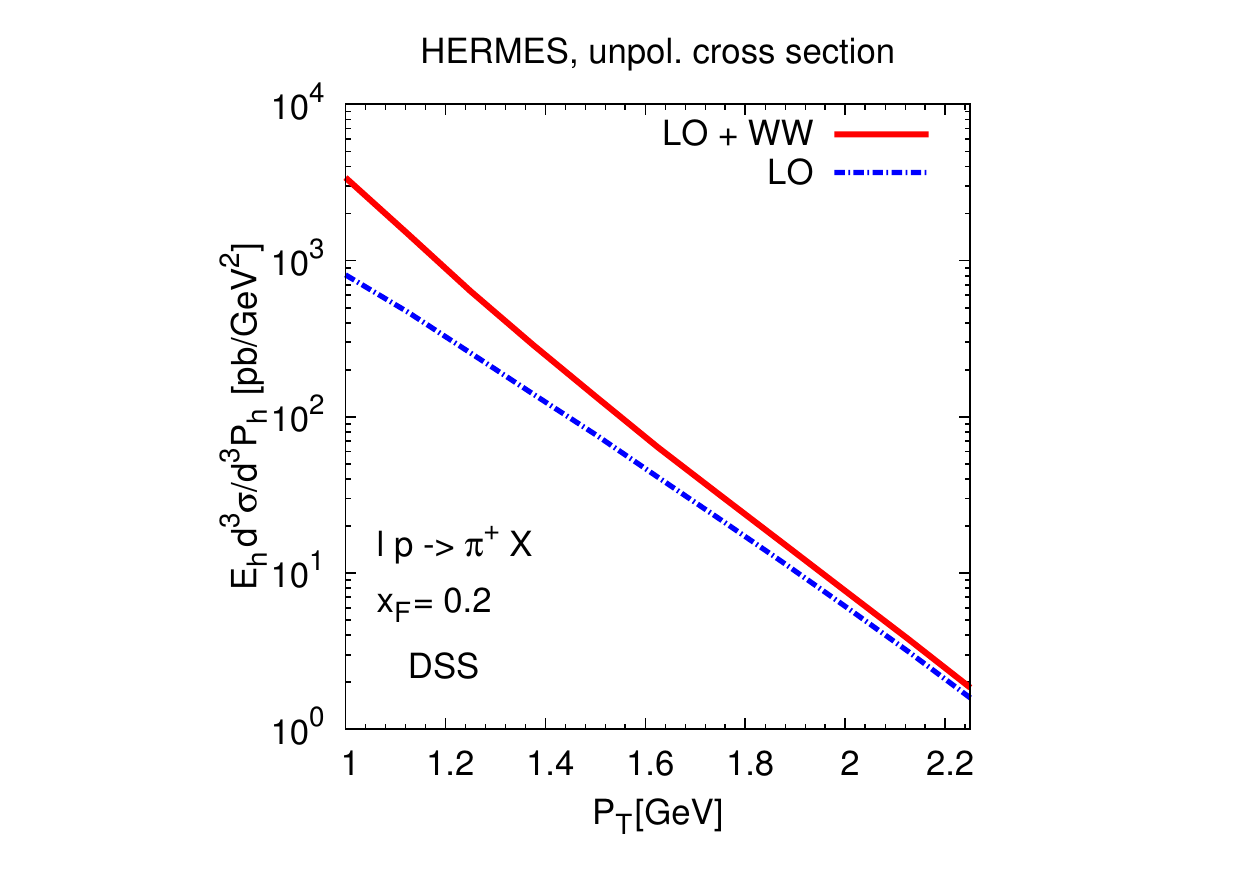}
 \caption{Unpolarized cross section at $x_F=0.2$ as a function of $P_T$ for $\ell p\to \pi^+\,X$, at HERMES, $\sqrt{s} = 7.25$ GeV, adopting two sets for the fragmentation functions: Kretzer set (left) and DSS set (right).}
  \label{fig:unp-herm-xF02}
\end{figure}

\begin{figure}[h!]
 \centering
 \includegraphics[scale=.65]{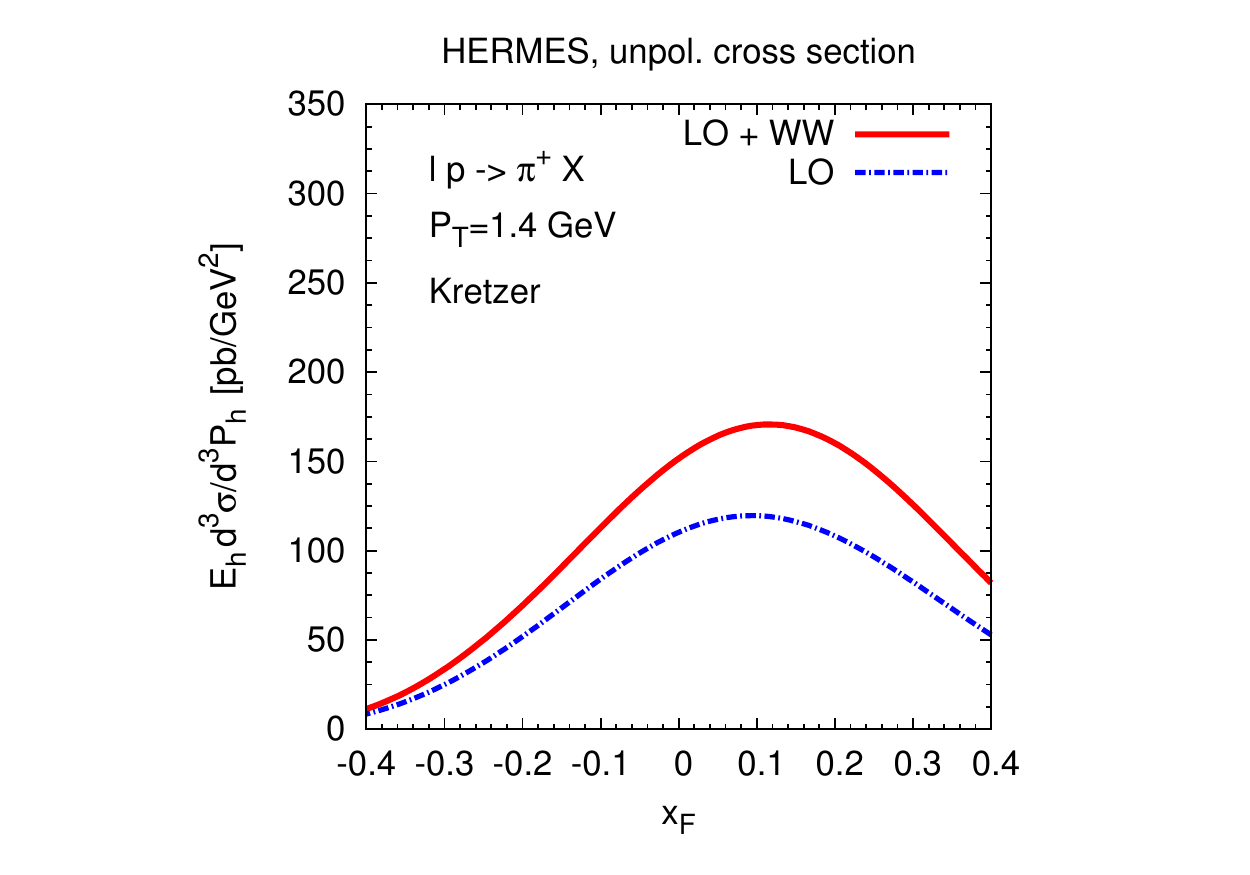}\hspace*{-2cm}
 \includegraphics[scale=.65]{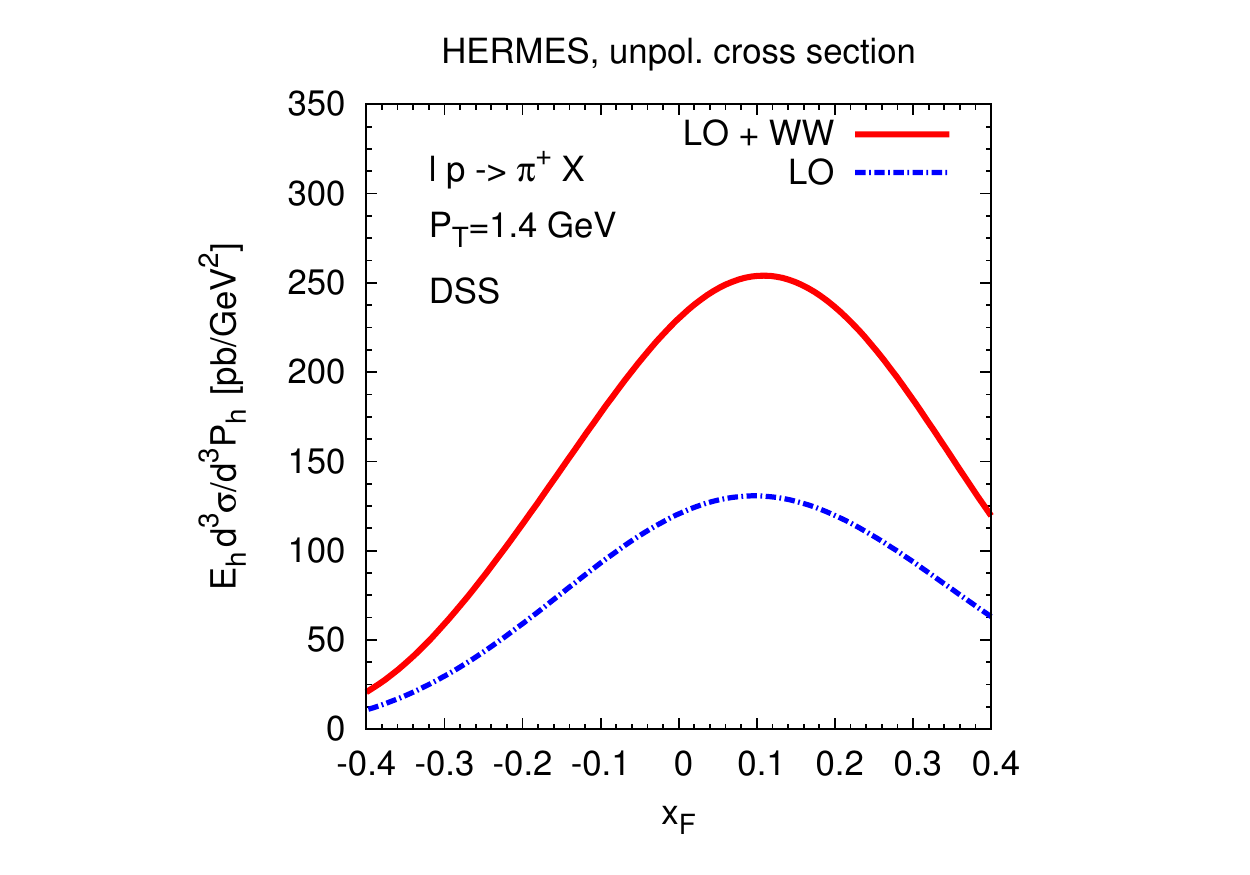}
 \caption{Unpolarized cross section at $P_T=1.4$ GeV as a function of $x_F$ for $\ell p\to \pi^+\,X$, at HERMES, $\sqrt{s} = 7.25$ GeV, adopting two sets for the fragmentation functions: Kretzer set (left) and DSS set (right).}
  \label{fig:unp-herm-pt14}
\end{figure}

In Figs.~\ref{fig:unp-herm-xF02} and \ref{fig:unp-herm-pt14} we present our estimates for the unpolarized cross sections for $\pi^+$ production at $\sqrt s\simeq 7.25$ GeV, respectively at fixed $x_F=0.2$ as a function of $P_T$, and at fixed $P_T=1.4$ GeV as a function of $x_F$. The blue dash-dotted lines represent the LO contribution, while the solid red lines include also the WW term. As one can see the quasi-real photon contribution is more important at lower $P_T$ values at fixed $x_F$ and is around 60-70\% of the total at fixed $P_T$. In particular, see Fig.~\ref{fig:unp-herm-pt14}, it is asymmetric in $x_F$, being more important for positive $x_F$ values. This is apparently counterintuitive, since for $x_F>0$ the lepton is scattered mainly in the backward region where one would expect a lesser role from quasi-real photon exchange.
On the other hand for large positive $x_F$, that for the HERMES set-up means that the pion is produced in the \emph{backward} proton hemisphere, $|\hat u| \ll |\hat t|$ and while the LO piece goes like $1/Q^2 \equiv 1/\hat t^2$, the partonic cross section for the subprocess $q\gamma\to q g$ (see Eq.~(\ref{M02}), first relation) goes like $1/\hat s\hat u$. Notice that the difference in size between the two computations, based on different FF sets, is due to the more important role of the gluon FF in the DSS set.


Let us now move to the predictions for the SSAs. In order to compare our estimates with HERMES data some further comments on the kinematical configurations and the notations adopted in their analysis are necessary. As already pointed out above in HERMES set-up~\cite{Airapetian:2013bim} the lepton is assumed to move along the positive $Z_{\rm cm}$ axis, so that the processes to be considered here are $\ell \, \pup \to h \, X$, rather than $\pup \ell \to h \, X$. In this reference frame the $\uparrow$ ($\downarrow$) direction is still along the $+Y_{\rm cm}$ ($-Y_{\rm cm}$) axis as in Ref.~\cite{Anselmino:2009pn} and, as we keep the usual definition of $x_F = 2 P_L/\sqrt s$, only its sign is reversed.

The azimuthal dependent cross section measured by HERMES is defined
as~\cite{Airapetian:2013bim}:
\be
d\sigma = d\sigma_{UU}[1+S_T \, A_{UT}^{\sin\psi} \sin\psi] \>,
\label{sigH}
\ee
where
\be
\sin \psi = \hat{\bm{S}}_T \cdot (\hat{\bm{P}}_T \times \hat{\bm{k}})
\ee
coincides with our $\sin\phi_S$ of Eq.~(\ref{phis}), as $\bm{p}$ and $\bm{k}$ (respectively, the proton and the lepton 3-momenta) are opposite vectors
in the lepton-proton {\it c.m.} frame. We then have
\be
A_{UT}^{\sin\psi}(x_F, P_T) = A_N^{p^\uparrow \ell \to h X}(-x_F, P_T)  \>,
\label{AUT-hermes}
\ee
where $A_N^{p^\uparrow \ell \to h X}$ is the SSA as given in Eqs.~(\ref{an}) and (\ref{ANWW}), and $A_{UT}^{\sin\psi}$ is the quantity measured by HERMES~\cite{Airapetian:2013bim}.

In the following we will consider both the fully-inclusive data as well as the sub-sample of anti-tagged data (with no detection of the final lepton) for $\ell \, p \to \pi \, X$ processes at large $P_T$.
In both cases there is only one large scale (needed for a perturbative calculation), the $P_T$ of the final pion. For this reason we only look at those data at $P_T\ge$ 1 GeV.

Notice that, at variance with SIDIS azimuthal asymmetries, one is not able to separate the single contributions to $A_N$ of the Sivers and Collins effects, that in principle could contribute together.

\begin{figure}[h!]
 \centering
 \includegraphics[scale=0.65]{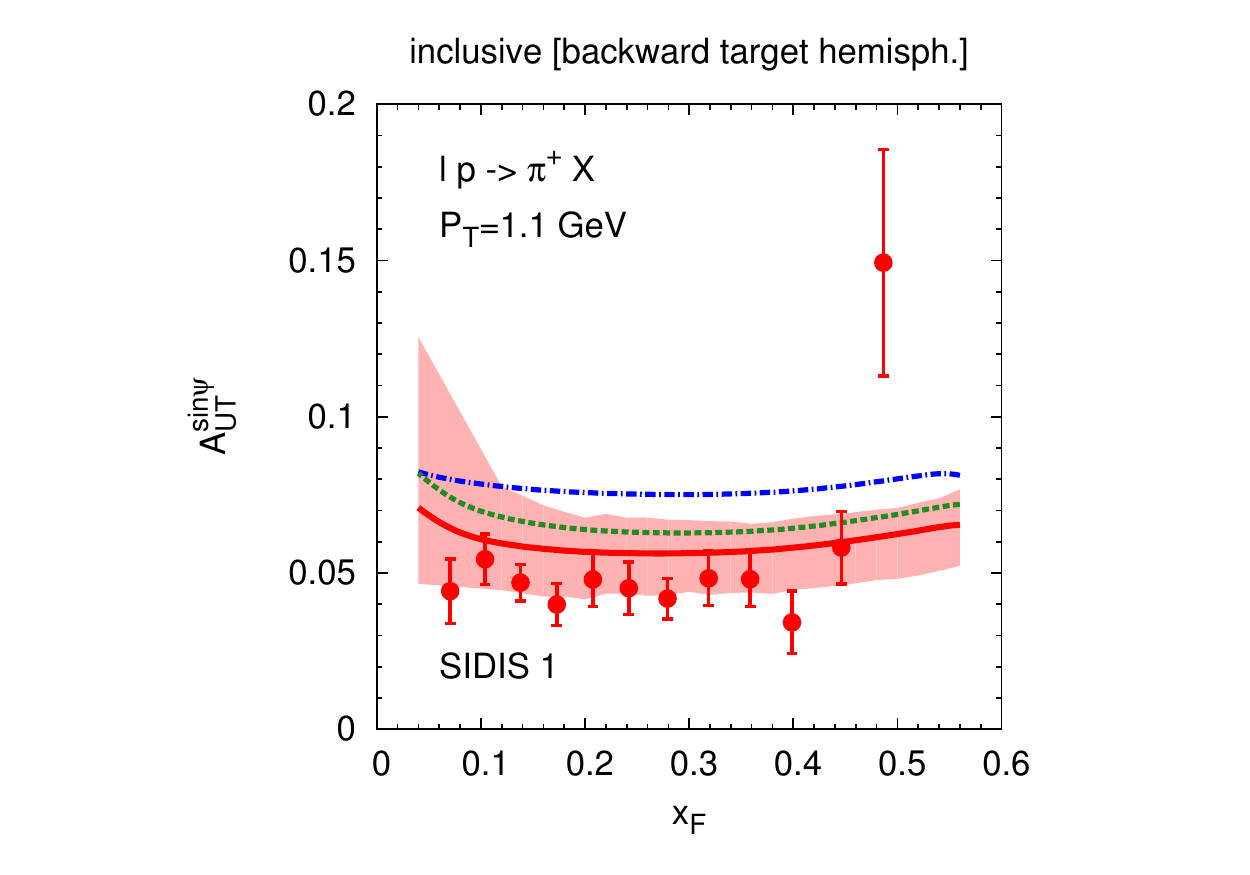}\hspace*{-1cm}
 \includegraphics[scale=0.65]{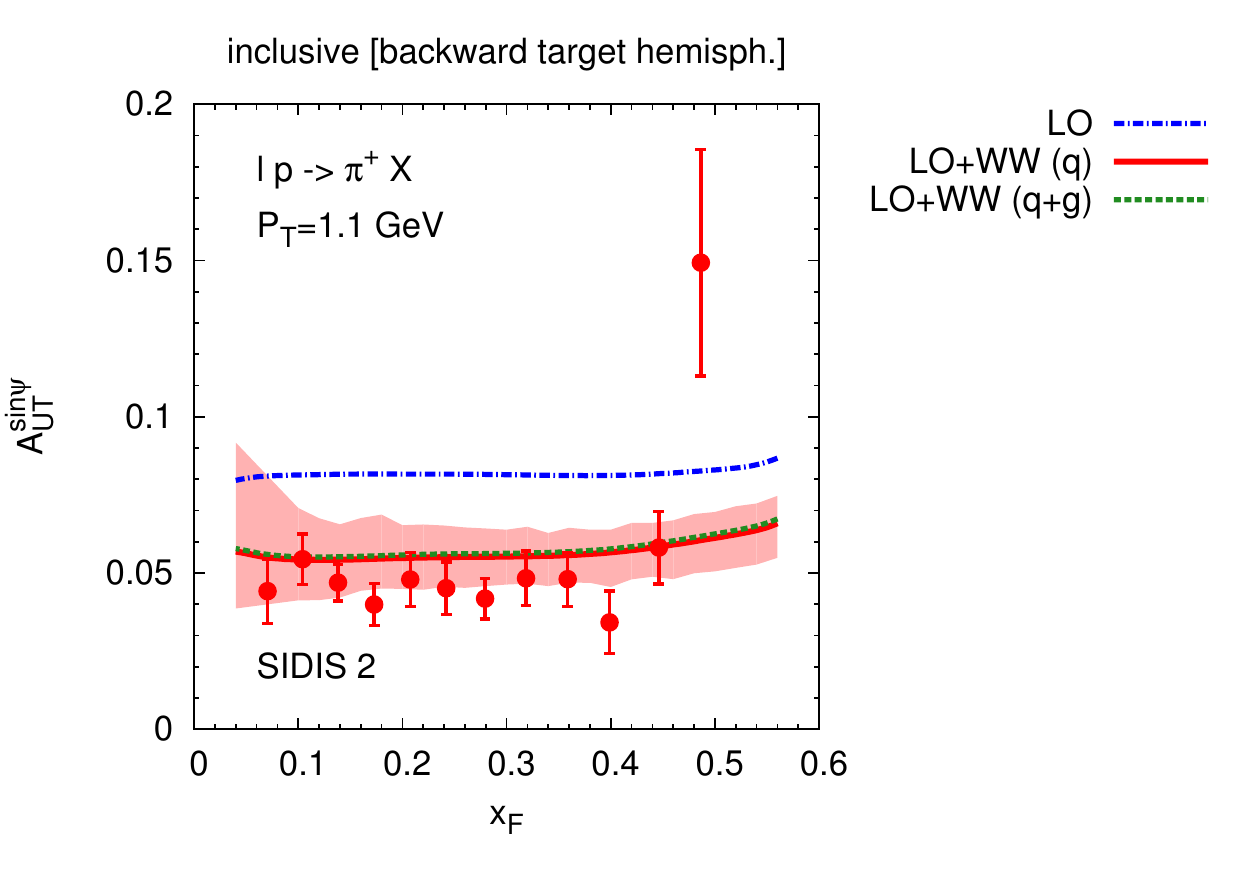}
  \includegraphics[scale=0.65]{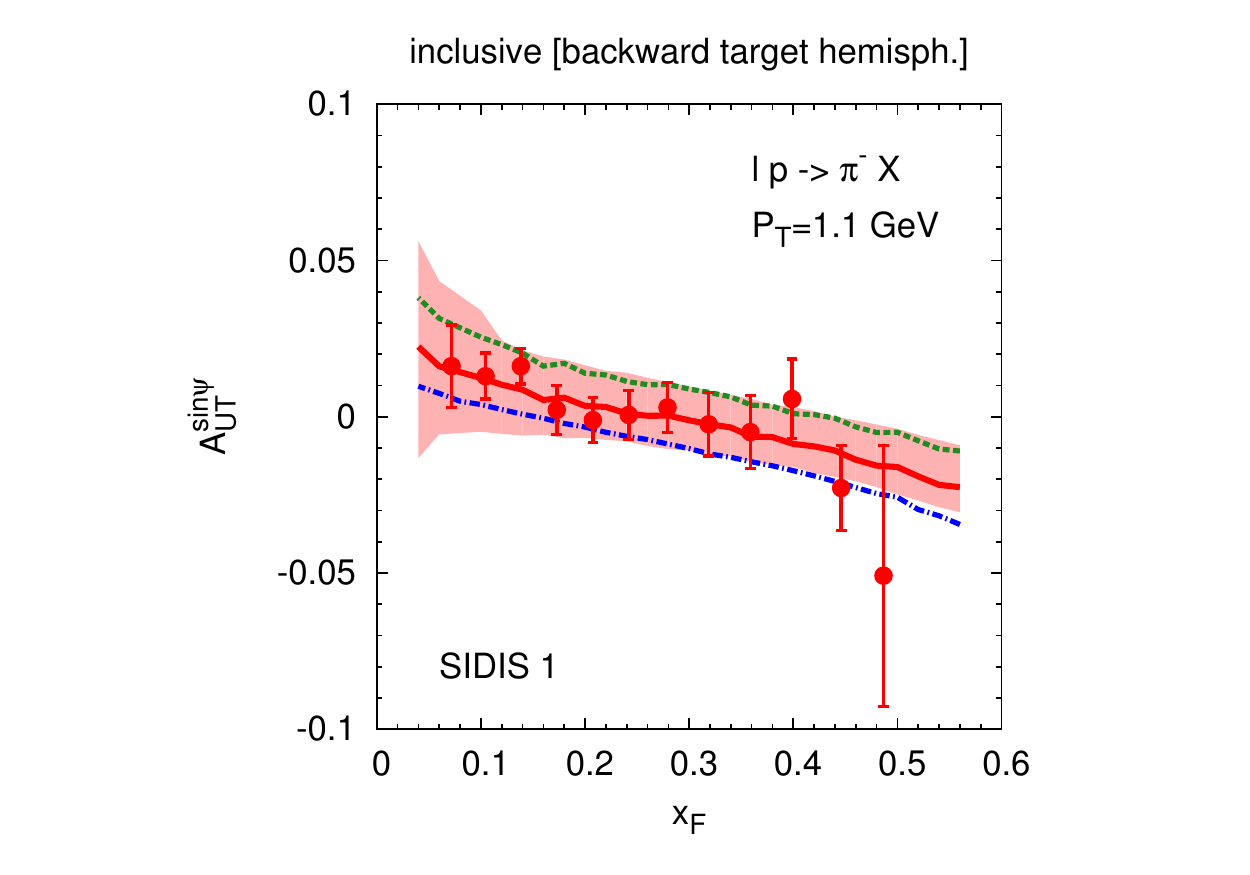}\hspace*{-1cm}
 \includegraphics[scale=0.65]{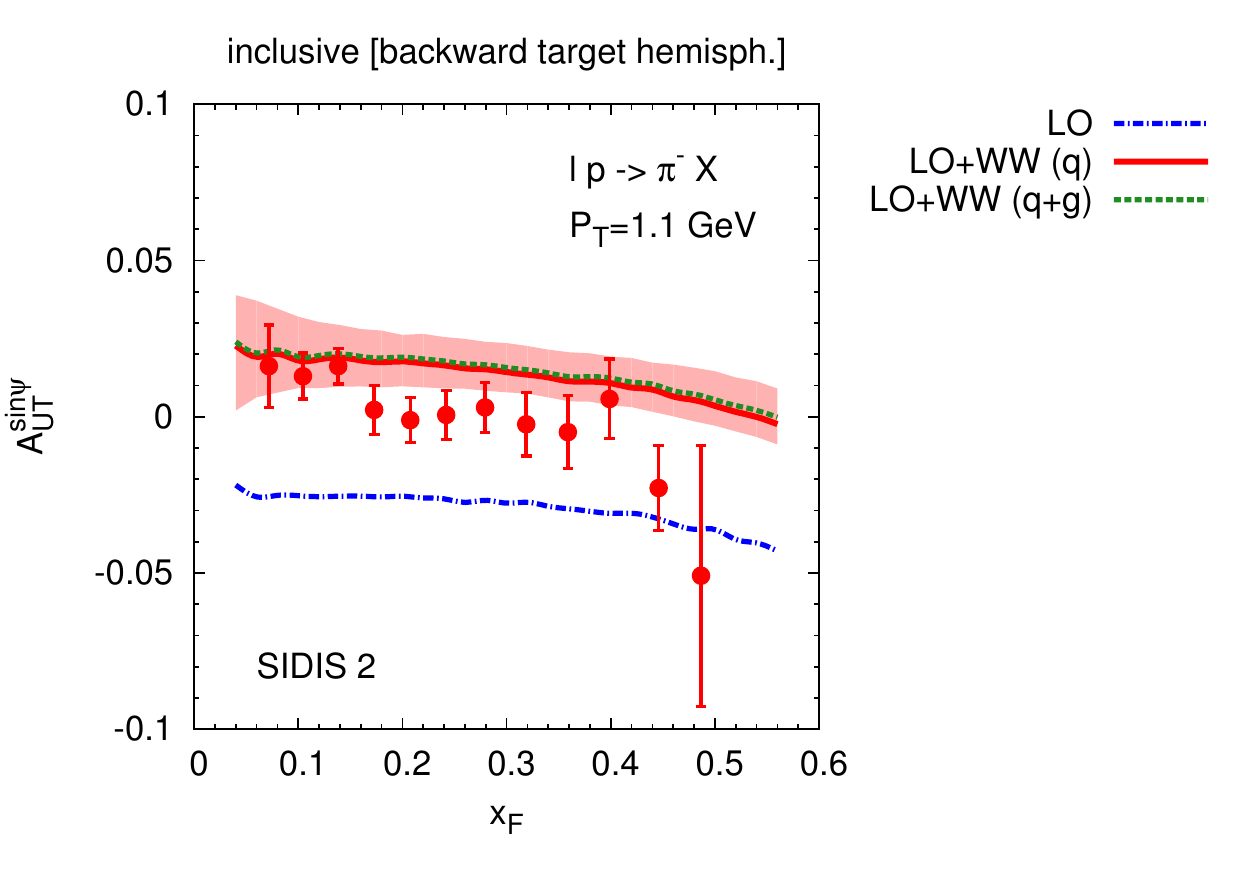}
\caption{The theoretical estimates for $A_{UT}^{\sin\psi}$ vs.~$x_F$ at $\sqrt{s}\simeq 7.25$ GeV and $P_T = 1.1$ GeV for inclusive $\pi^+$ (upper panels) and $\pi^-$ (lower panels) production in $\ell \, \pup \to \pi \, X$ processes, compared with the HERMES data~\cite{Airapetian:2013bim}. See text and legend for details.}
 \label{fig:SSA-pi-pt11}
\end{figure}

\begin{figure}[h!]
 \centering
 \includegraphics[scale=0.65]{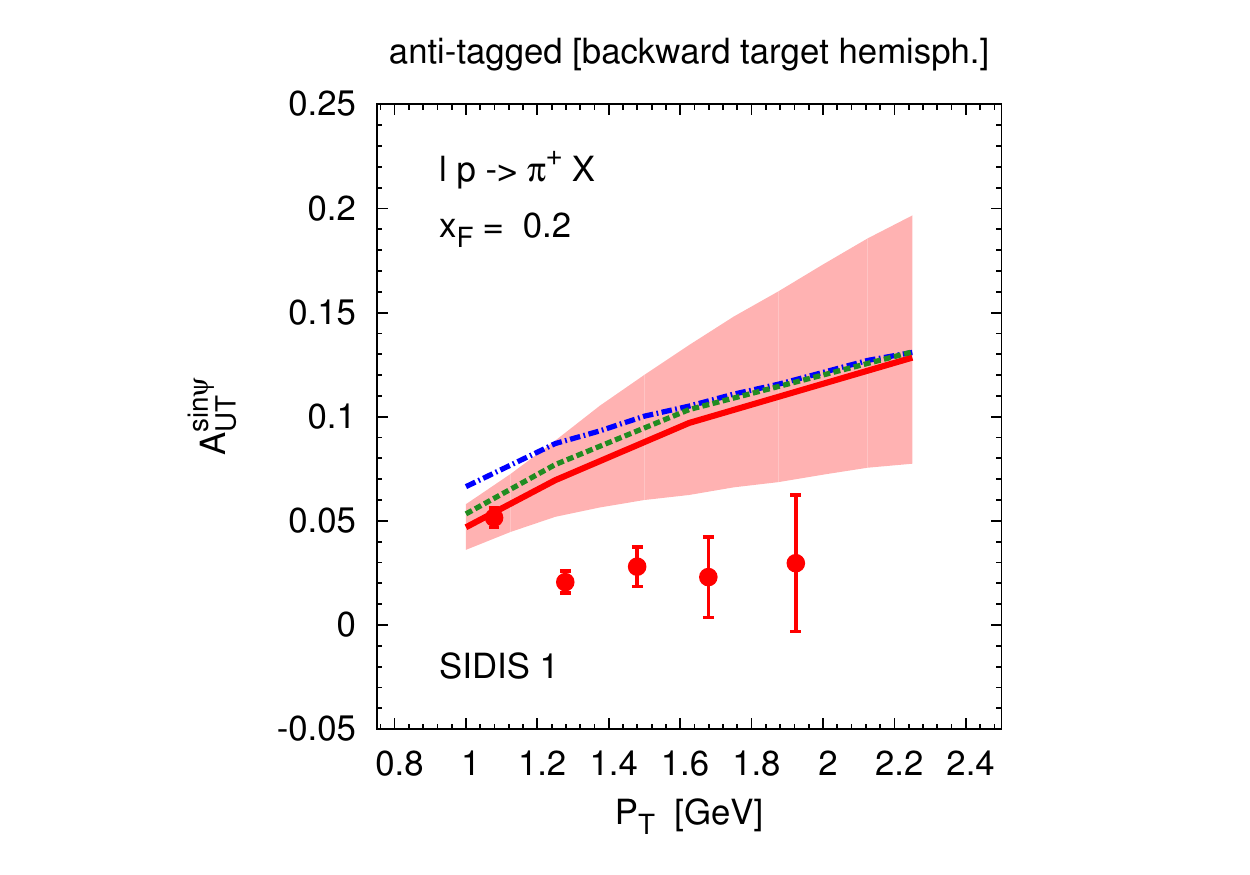}\hspace*{-1cm}
 \includegraphics[scale=0.65]{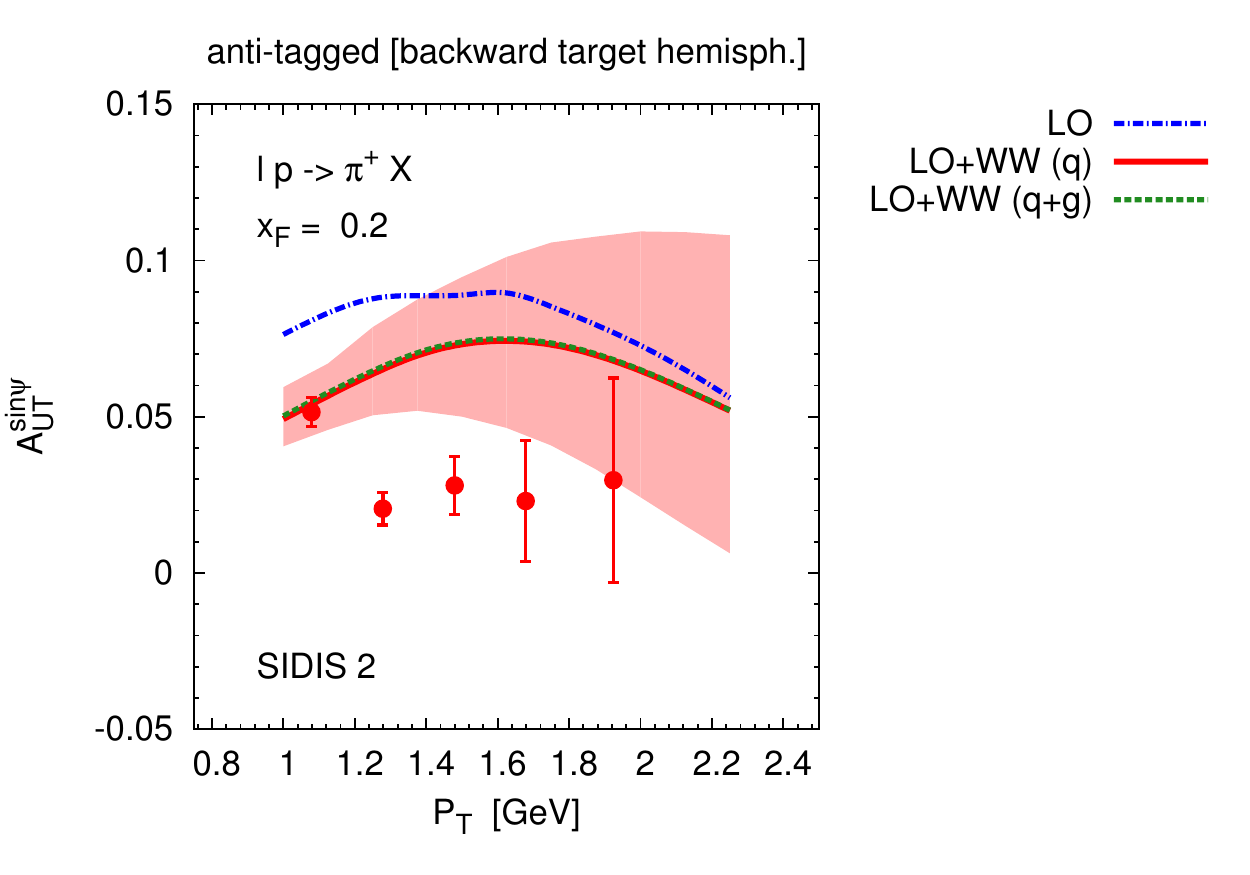}
  \includegraphics[scale=0.65]{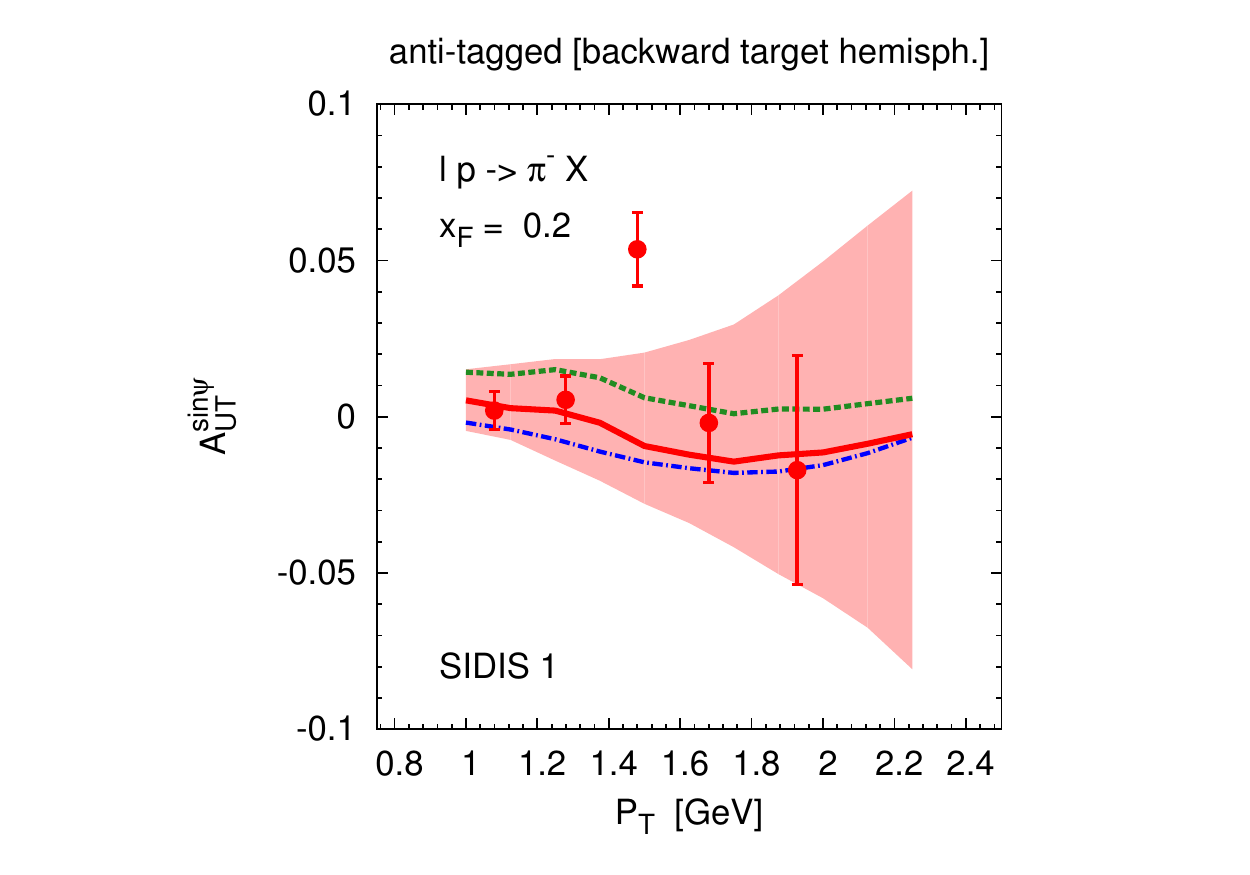}\hspace*{-1cm}
 \includegraphics[scale=0.65]{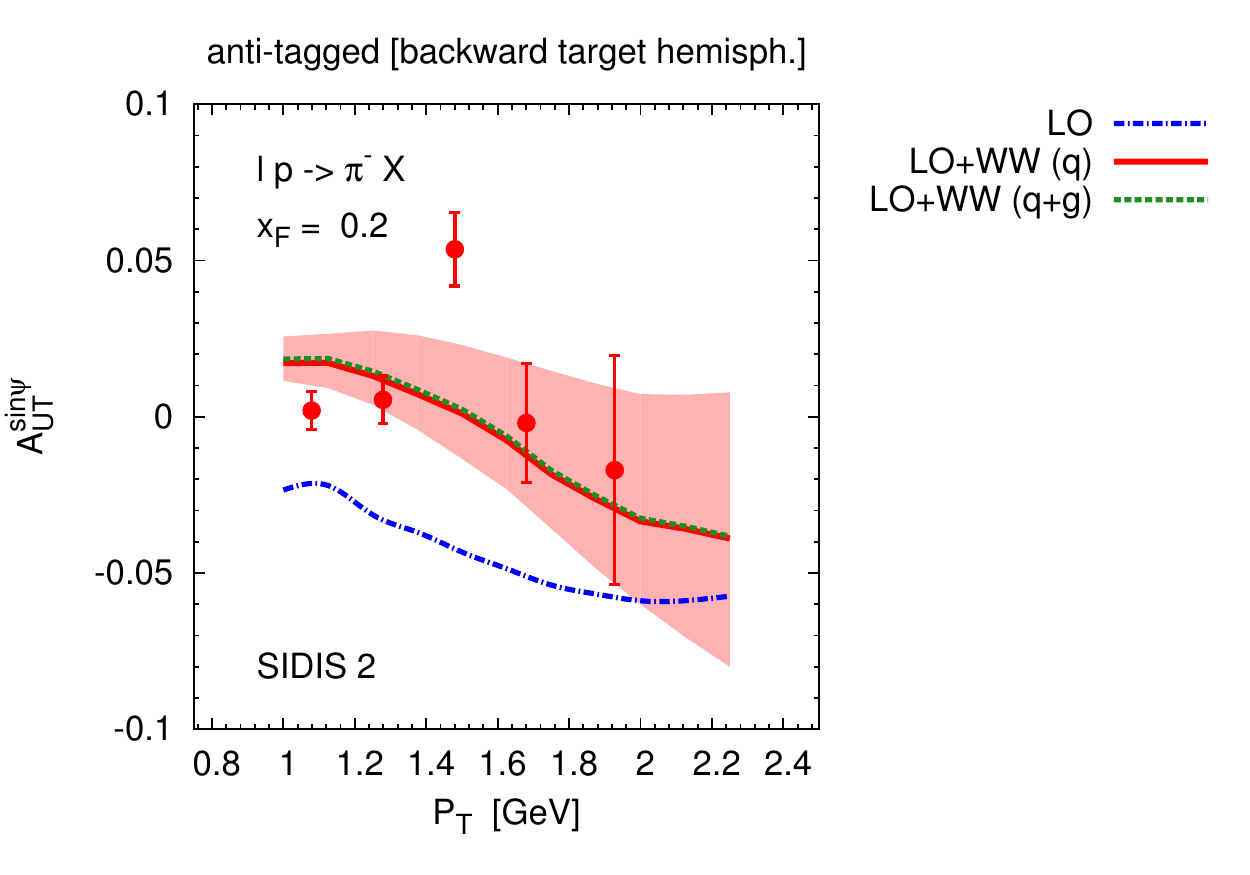}
\caption {The theoretical estimates for $A_{UT}^{\sin\psi}$ vs.~$P_T$ at $\sqrt{s}\simeq 7.25$ GeV and $x_F = 0.2$ for inclusive $\pi^+$ (upper panels) and $\pi^-$ (lower panels) production in $\ell \, \pup \to \pi \, X$ processes, compared with the HERMES anti-tagged data~\cite{Airapetian:2013bim}. See text and legend for details.}
 \label{fig:SSA-pi-xf02}
\end{figure}

Our predictions for $A_{UT}^{\sin\psi}$ compared with HERMES data are shown in Figs.~\ref{fig:SSA-pi-pt11} (inclusive data set, vs.~$x_F$ at $P_T = 1.1$ GeV) and \ref{fig:SSA-pi-xf02} (anti-tagged data set, vs.~$P_T$ at $x_F = 0.2$). More precisely, we show the quark Sivers and Collins contributions, adopting the SIDIS 1 (left panels) and SIDIS 2 (right panels) sets,  at LO (blue dot-dashed lines) and at LO + WW (red solid lines), as well as the total result adding the contribution from the gluon Sivers function (green dotted lines). The overall statistical uncertainty band, also shown, is the envelope of the independent statistical uncertainty bands on the Sivers and Collins functions for quarks, obtained following the procedure described in Appendix A of Ref.~\cite{Anselmino:2008sga}.

Some comments are in order. Let us start with the fully-inclusive case, Fig.~\ref{fig:SSA-pi-pt11}: the inclusion of the WW contribution improves significantly the agreement with the data (remember that in this kinematical region it is the dominant part in the unpolarized cross sections); the Collins effect is always tiny or negligible (both in the LO and WW contributions); the differences between the predictions adopting the SIDIS~1 and the SIDIS~2 sets are due to the different behaviour of the corresponding Sivers functions; the contribution coming from the gluon Sivers function is almost negligible for the SIDIS~2 set, while for the SIDIS~1 set is relatively more important, reducing the agreement with the data\footnote{Notice that there is  still a large uncertainty in the gluon Sivers function extraction in the large-$x$ region.}. Concerning the anti-tagged data set, Fig.~\ref{fig:SSA-pi-xf02}, we could observe that: once again the WW contribution leads to a very good description of the data (even if some sizeable discrepancy for the $\pi^+$ data remains). The gluon Sivers effect is negligible, except for the SIDIS~1 set in $\pi^-$ production. However, one has to keep in mind that this kinematical region probes the still poorly constrained large-$x$ behaviour of the Sivers functions (the dominant contribution). This is the reason of the wider error bands.
Quite interestingly a very recent extraction of the Sivers functions from SIDIS data seems to reduce significantly the discrepancies between the theoretical predictions and the anti-tagged data set, see Ref.~\cite{DAlesio:2017nrd}.

\section{Conclusions}
The issue of TMD factorization and universality is crucial for our understanding of SSAs in QCD. Here we have further pursued the idea, already discussed in a previous study, for assessing the validity of a TMD scheme in inclusive processes in which a single large-$P_T$ particle is produced. We have computed the single spin asymmetry $A_N$, for the $\ell \, \pup \to h \, X$ process, as generated by the
Sivers and the Collins functions, which have been extracted from SIDIS and $e^+e^-$ data~\cite{Anselmino:2005ea,Anselmino:2007fs,Anselmino:2008sga,
Anselmino:2008jk}. Doing so, we adopt a unified TMD factorized approach, valid for $\ell \, p \to \ell \, h \, X$ and $\ell \, p \to h \, X$
processes, in which, consistently, we obtain information on the TMDs and make predictions for $A_N$.

In the present analysis we have extended this strategy, including the contribution of quasi-real photon exchange, expected to be important when the final lepton is scattered at small angles. We have indeed obtained that for HERMES kinematics the WW contribution to the unpolarized cross section is huge, reaching more than 70\%, and therefore dominating over the LO term. Moreover, and more importantly, we have shown how the data description of the SSAs observed at HERMES, concerning their size, sign and behaviour, (already quite satisfying at LO) is definitely improved when one includes the WW piece. This seems corroborating the overall approach.\\

U.D.~thanks the organizers for their kind invitation to such a nice and fruitful workshop.


\begin{thebibliography}{99}

\bibitem{D'Alesio:2007jt}
  U.~D'Alesio and F.~Murgia,
  Prog.\ Part.\ Nucl.\ Phys.\  {\bf 61} (2008) 394
  [arXiv:0712.4328].



\bibitem{Anselmino:2009pn}
  M.~Anselmino, M.~Boglione, U.~D'Alesio, S.~Melis, F.~Murgia and A.~Prokudin,
  Phys.\ Rev.\ D {\bf 81} (2010) 034007
  [arXiv:0911.1744].



\bibitem{Anselmino:2014eza}
  M.~Anselmino, M.~Boglione, U.~D'Alesio, S.~Melis, F.~Murgia and A.~Prokudin,
  Phys.\ Rev.\ D {\bf 89} (2014),  114026
  [arXiv:1404.6465].



\bibitem{Collins:2011zzd}
  J.~Collins, \emph{Foundations of Perturbative QCD}, Cambridge Monographs on Particle Physics, Nuclear
  Physics and Cosmology, Vol. 32 (Cambridge University Press, Cambridge, 2011).



\bibitem{GarciaEchevarria:2011rb}
  M.~G.~Echevarria, A.~Idilbi and I.~Scimemi,
  JHEP {\bf 1207} (2012) 002
  [arXiv:1111.4996].



\bibitem{Echevarria:2014rua}
  M.~G.~Echevarria, A.~Idilbi and I.~Scimemi,
  Phys.\ Rev.\ D {\bf 90} (2014),  014003
  [arXiv:1402.0869].


\bibitem{Koike:2002gm}
  Y.~Koike,
  Nucl.\ Phys.\ A {\bf 721} (2003) 364
  [hep-ph/0211400].

\bibitem{Gamberg:2014eia}
  L.~Gamberg, Z.~B.~Kang, A.~Metz, D.~Pitonyak and A.~Prokudin,
  Phys.\ Rev.\ D {\bf 90} (2014),  074012
  [arXiv:1407.5078].



\bibitem{Airapetian:2013bim}
  A.~Airapetian {\it et al.} [HERMES Collaboration],
  Phys.\ Lett.\ B {\bf 728} (2014) 183
  [arXiv:1310.5070].

\bibitem{DAlesio:2017nrd}
  Umberto D'Alesio, Carlo Flore, and Francesco Murgia, [arXiv:1701.01148].

\bibitem{Hinderer:2015hra}
  P.~Hinderer, M.~Schlegel and W.~Vogelsang,
  Phys.\ Rev.\ D {\bf 92} (2015),  014001;
   Erratum: [Phys.\ Rev.\ D {\bf 93} (2016),  119903]
  [arXiv:1505.06415].

\bibitem{Sivers:1989cc}
  D.~W.~Sivers,
  Phys.\ Rev.\ D {\bf 41} (1990) 83.

\bibitem{Sivers:1990fh}
  D.~W.~Sivers,
  Phys.\ Rev.\ D {\bf 43} (1991) 261.

\bibitem{Collins:1992kk}
  J.~C.~Collins,
  Nucl.\ Phys.\ B {\bf 396} (1993) 161
  [hep-ph/9208213].

\bibitem{D'Alesio:2004up}
  U.~D'Alesio and F.~Murgia,
  Phys.\ Rev.\ D {\bf 70} (2004) 074009
  [hep-ph/0408092].



\bibitem{Anselmino:2005sh}
  M.~Anselmino, M.~Boglione, U.~D'Alesio, E.~Leader, S.~Melis and F.~Murgia,
  Phys.\ Rev.\ D {\bf 73} (2006) 014020
  [hep-ph/0509035].




\bibitem{Anselmino:2005ea}
  M.~Anselmino, M.~Boglione, U.~D'Alesio, A.~Kotzinian, F.~Murgia and A.~Prokudin,
  Phys.\ Rev.\ D {\bf 72} (2005) 094007;
   Erratum: [Phys.\ Rev.\ D {\bf 72} (2005) 099903]
  [hep-ph/0507181].



\bibitem{Anselmino:2007fs}
  M.~Anselmino, M.~Boglione, U.~D'Alesio, A.~Kotzinian, F.~Murgia, A.~Prokudin and C.~T\"urk,
  Phys.\ Rev.\ D {\bf 75} (2007) 054032
  [hep-ph/0701006].



\bibitem{Anselmino:2008sga}
  M.~Anselmino, M.~Boglione, U.~D'Alesio, A.~Kotzinian, S.~Melis, F.~Murgia, A.~Prokudin and C.~T\"urk,
  Eur.\ Phys.\ J.\ A {\bf 39} (2009) 89
  [arXiv:0805.2677].



\bibitem{Anselmino:2008jk}
  M.~Anselmino, M.~Boglione, U.~D'Alesio, A.~Kotzinian, F.~Murgia, A.~Prokudin and S.~Melis,
  Nucl.\ Phys.\ Proc.\ Suppl.\  {\bf 191} (2009) 98
  [arXiv:0812.4366].



\bibitem{Kretzer:2000yf}
  S.~Kretzer,
  Phys.\ Rev.\ D {\bf 62} (2000) 054001
  [hep-ph/0003177].



\bibitem{deFlorian:2007aj}
  D.~de Florian, R.~Sassot and M.~Stratmann,
  Phys.\ Rev.\ D {\bf 75} (2007) 114010
  [hep-ph/0703242].



\bibitem{D'Alesio:2015uta}
  U.~D'Alesio, F.~Murgia and C.~Pisano,
  JHEP {\bf 1509} (2015) 119
  [arXiv:1506.03078].


\end{thebibliography}
\end{document}